\documentclass{acm_proc_article-sp}

\usepackage{amsmath}
\usepackage{textcomp}
\usepackage{marvosym}
\usepackage{wasysym}
\usepackage{amssymb}
\usepackage{hyperref}

\usepackage{enumitem}
\setlist{noitemsep, topsep=0pt}
\let\oldenumerate\enumerate
\renewcommand{\enumerate}{\vspace{-\topsep}\oldenumerate}

\usepackage{listings}
\usepackage{caption}
\begin{document}

\newcommand{\eclipse}{ECL\textsuperscript{i}PS\textsuperscript{e}}

\makeatletter
\newcommand*{\bdiv}{%
  \nonscript\mskip-\medmuskip\mkern5mu%
  \mathbin{\operator@font div}\penalty900\mkern5mu%
  \nonscript\mskip-\medmuskip
}
\makeatother

\title{Declaratively solving tricky Google Code Jam problems with Prolog-based ECLiPSe CLP system}

\numberofauthors{2}
\author{
\alignauthor
Sergii Dymchenko\\
       \affaddr{Independent Researcher}\\
       \email{sdymchenko@progopedia.com}
\alignauthor
Mariia Mykhailova\\
       \affaddr{Independent Researcher}\\
       \email{michaylova@gmail.com}
}
\maketitle

\begin{abstract}
In this paper we demonstrate several examples of solving challenging algorithmic problems from the Google Code Jam programming contest
with the Prolog-based \eclipse\ system using declarative techniques: constraint logic programming and linear (integer) programming.
These problems were designed to be solved by inventing clever algorithms and efficiently implementing them in a conventional imperative programming language,
but we present relatively simple declarative programs in \eclipse\ that are fast enough to find answers within the time limit imposed by the contest rules.
We claim that declarative programming with \eclipse\ is better suited for solving certain common kinds of programming problems offered in Google Code Jam than imperative programming.
We show this by comparing the mental steps required to come up with both kinds of solutions.

\end{abstract}

\category{D.3.2}{Programming Languages}{Language Classifications}[Constraint and logic languages]
\category{G.1.6}{Numerical Analysis}{Optimization}[Linear programming]

\keywords{Declarative programming, logic programming, constraint programming, linear programming, programming competitions}

\section{Introduction}
Google Code Jam\footnote{\url{https://code.google.com/codejam}} (GCJ) is one of the
biggest programming competitions in the world: almost 50,000 participants registered in 2014, and 25,462 of them solved at least one task.

For good results, competitors must think and code quickly -- an individual round is usually 2--4 hours long and poses 3 or more problems. 
A solution is considered correct if it produces correct answers for all given test cases within a certain time limit 
(4 minutes for the ``small" input and 8 minutes for the ``large" one).

GCJ competitors can use any freely available programming language or system (including the \eclipse\footnote{\url{http://www.eclipseclp.org/}} system described in this paper). 
Most other competitions restrict participants to a limited set of popular programming languages (typically C++, Java, C\#, Python).
Many contestants participate not only in GCJ, but also in other contests, and use the same language across all competitions.
When the GCJ problem setters design the problems and evaluate their complexity, they keep in mind mostly this crowd of seasoned algorithmists. 

We show that some GCJ problems that should be challenging according to the problem setters' estimate 
can be easy or even trivial to 
solve declaratively using high-level programming languages like \eclipse\ and techniques like constraint logic programming and linear (integer) programming.

\eclipse\ CLP\cite{schimpf2012eclipse,apt2007constraint}
is an open-source software system which aims to serve as a platform for integrating various logic programming extensions, in particular constraint logic programming (CLP).
\eclipse\ has a modern and efficient implementation of the Prolog programming language in its core, offers some Prolog extensions (like declarative loops \cite{schimpf2002loops}),
contains several constraint programming libraries (`ic' for interval arithmetic, `fd' for finite domains, `grasper' for graphs), and
interfaces to third-party solvers (Gecode, Gurobi, COIN-OR solvers, CPLEX) \cite{eclipse-library-manual}

\subsection*{\eclipse\ implementation of TPK algorithm}

We do not assume that the reader is familiar with \eclipse, so to give a feeling of the syntax we present and explain a program for a TPK algorithm.

TPK is a simple algorithm proposed by D.\,E.\,Knuth and L.\,T.\,Pardo  \cite{knuth1976early}. 
It is used to demonstrate some basic syntactic constructs of a programming language beyond the ``Hello, World!'' program.
The algorithm prompts for 11 real numbers ($a_0 \dots a_{10}$) and for each $a_i$ computes $b_i = f(a_i)$, where $f(t) = \sqrt{|t|} + 5 {t}^3$.
After that for $i = 10 \dots 0$ (in that order) the algorithm outputs a pair $(i, b_i)$ if $b_i \leq 400$, or $(i,$ TOO LARGE$)$ otherwise.
\clearpage
\begin{lstlisting}[caption={TPK algorithm in \eclipse}]
f(T, Y) :-
    Y is sqrt(abs(T)) + 5*T^3.
main :-
    read(As), 
    length(As, N), reverse(As, Rs),
    ( foreach(Ai, Rs), for(I, N - 1, 0, -1) do
        Bi is f(Ai),
        ( Bi > 400 ->
            printf("%w TOO LARGE\n", I)
        ;
            printf("%w %w\n", [I, Bi])
        )
    ).
\end{lstlisting}

Lines 1--2 define a predicate to calculate value of $f$. 
Lines 3--13 define predicate \texttt{main}.  
The name of this predicate is passed to \eclipse\ translator as a command-line parameter.

Line 4 reads a list into variable \texttt{As}.
The input is a Prolog list (comma-delimited and bracket-enclosed).
We preprocess the space-separated input with a simple `sed' script to convert it to the Prolog format.

Line 5 stores the length of the input list in the variable \texttt{N} (the program can work with inputs of different lengths, not necessarily 11 as original TPK program),
and stores original input numbers in reverse order in the variable \texttt{Rs}.

Line 6 is the head of \eclipse\ loop: we iterate simultaneously over every number in \texttt{Rs} and over $I = N-1 \dots 0$. 
\texttt{I}, \texttt{Ai} and \texttt{Bi} are local variables for every loop iteration.
Line 7 simply assigns the value \texttt{f(Ai)} to \texttt{Bi}. 
This line demonstrates \eclipse\ support of using arithmetic predicates as functions -- in many other Prolog implementations less clear \texttt{f(Ai, Bi)} must be used.
Lines 8--10 are Prolog `if-then-else' construct that outputs ``TOO LARGE'' or \texttt{Bi} depending of the value of \texttt{Bi}.
\texttt{printf} is very similar to the corresponding C function. \texttt{\%w} is a ``wildcard'' control sequence that can be used to output values of different types.
The second argument of \texttt{printf} is a value or a list of values for substitution.

Indentation has no syntactic meaning in Prolog (or \eclipse), we use it for clarity.

\section{The Problems}

In this section we show how tricky algorithmic problems can often be easily modeled and efficiently solved using declarative programming techniques and \eclipse.
We chose a set of GCJ problems from different tournament stages to demonstrate different useful aspects of \eclipse: constraint programming library `ic', 
linear programming library `eplex', working with integers and floating point numbers.

To compare our declarative programming solutions with possible imperative solutions we compare the ``mental steps'' required to come up with the solution
(counting number of lines of code would not be useful because the main challenge is to invent the solution, not to code it.) 
Our \eclipse\ programs solve both large and small inputs for each problem.

\subsection*{Triangle Areas\footnote{Problem link: \url{http://goo.gl/enHWlq}}}

``Triangle Areas'' is a problem from the second round of GCJ 2008.
The problem statement can be rephrased as follows: given integer $N$, $M$ and $A$, find a
triangle with vertices in integer points with coordinates $0 \leq x_i \leq N$
and $0 \leq y_i \leq M$ that has an area equal to $\frac{A}{2}$, or say that it does not exist.

``Triangle Areas'' is almost perfect for solving with constraint logic programming. 
Variables are discrete, constraints are non-linear (so integer linear programming can not be used), and we are looking for any feasible solution.
The problem is not too hard from purely algorithmic point of view, but correct implementation using conventional programming languages is very tricky.
Many contestants who solved all other problems (including higher valued problems) failed to solve the large or both inputs for ``Triangle Areas".
At the same time, modeling this problem in \eclipse\ is almost trivial. 
To come up with an effective model we need to notice that one vertex of the triangle can be chosen arbitrarily.
With this observation, the most convenient way to calculate the doubled triangle area is to place one vertex in $(0,0)$; then $2S = |x_2y_3 - x_3y_2|$. 
(The same formula can be used in an imperative solution.)

For this problem we present complete source code of the solution. 
For subsequent problems we present only interesting parts: `model' plus other problem-specific predicates.

\begin{lstlisting}[caption={Complete \eclipse\ program for the ``Triangle Areas'' problem}]
:- lib(ic).
model(N, M, A, [X2, Y2, X3, Y3]) :-
    [X2, X3] :: 0..N,
    [Y2, Y3] :: 0..M,
    A #= abs(X2 * Y3 - X3 * Y2).
do_case(Case_num, N, M, A) :-
    printf("Case #%w: ", [Case_num]),
    ( model(N, M, A, Points), labeling(Points) ->
        printf("0 0 %w %w %w %w", Points)
    ; 
        write("IMPOSSIBLE") 
    ),
    nl.
main :-
    read([C]), 
    ( for(Case_num, 1, C) do 
        read([N, M, A]),
        do_case(Case_num, N, M, A) ).
\end{lstlisting}

Line 1 loads interval arithmetic constraint programming library `ic'. 
Lines 2--5 define the model with input parameters \texttt{N}, \texttt{M}, \texttt{A} and a list of output parameters \texttt{[X2, Y2, X3, Y3]}.
\texttt{::} and \texttt{\#=} are from `ic' library. 
With \texttt{::} we define possible domains for \texttt{X2, X3, Y2, Y3} variables,
and \texttt{\#=} from `ic' constraints both left and right parts to be equal and integer.
After model evaluation \texttt{X2, X3, Y2, Y3} variables will not necessarily be instantiated to concrete values, but they will have reduced domains with possible delayed constraints 
and will be instantiated later with \texttt{labeling}.

Lines 6--13 define the \texttt{do\_case} predicate to process a single input case. Line 7 outputs case number according to the problem specification.
Lines 8--12 are an `if-then-else' construct that outputs point coordinates if it is possible to satisfy our \texttt{model} predicate and labels (assigns concrete values from the domain)
all coordinate variables, or ``IMPOSSIBLE'' otherwise.
Line 13 simply outputs a new line character.

Lines 14--18 define the \texttt{main} predicate that reads number of test cases \texttt{C} and for each test case reads \texttt{N, M, A} parameters and executes \texttt{do\_case}.

We can formulate following mental steps needed to invent and implement this \eclipse\ solution:

\begin{enumerate}
\item \textit{Notice that one vertex can be chosen as $(0,0)$.}
\item \textit{Recall or look up the formula for an area of a triangle.}
\item \textit{Formulate the constraint programming model.}
\item \textit{Code the constraint programming model.} 
For this problem, it requires 4 lines of straightforward code.
\end{enumerate}

Let us compare this with a possible imperative
solution in a convenient programming language that requires a more in-depth analysis of the problem.
First observations will be the same. We will also note that it is impossible to find required triangle
if $A > M \times N$, and for $A = M \times N$ triangle $(0,0), (N,0), (0,M)$ is a valid answer.
Now, for $A < M \times N$ we can represent A as $M(A \bdiv M) + (A \bmod M)$, $0 < A \bdiv M < N, 0 < A \bmod M < M$. 
If we match this representation with the area formula, we can see that points $(0, 0), (1, M)$, and $(- A \bdiv M, A \bmod M)$
form a triangle with area $\frac{A}{2}$. If we shift this triangle $A \bdiv M$ units in positive direction along the $x$ axis, we will get 
a triangle $(A \bdiv M, 0), (A \bdiv M + 1, M), (0, A \bmod M)$ that will match all the requirements.

The mental steps for this solution could be:

\begin{enumerate}
\item \textit{Notice that one vertex can be chosen as $(0,0)$.}
\item \textit{Recall or look up the formula for an area of a triangle.}
\item \textit{Figure out that for $A > M \times N$ there is no such triangle.}
\item \textit{Find the solution for border case $A = M \times N$.}
\item \textit{Come up with a representation of A that matches the area formula for a special triangle.}
\item \textit{Shift this triangle to fit inside the boundaries.}
\item \textit{Code the solution.} The code is even easier, check how $A$ relates to $M \times N$ and output corresponding triangle.
\end{enumerate}

Arguably, declarative solution in \eclipse\ needs simpler steps and leaves less space for a possible mistake.

\subsection*{Dancing With the Googlers\footnote{Problem link: \url{http://goo.gl/JpQQYi}}}

``Dancing With the Googlers" is a problem from the qualification round of GCJ 2012.
In this problem we consider triplets of integers from 0 to 10 which
never contain numbers that are more than 2 apart. A triplet is
surprising if it contains numbers that are exactly 2 apart. Given the list of
sums of $N$ triplets and the number of surprising triplets among them
$S$, how many triplets can be high (have the highest number at least $p$)?
 
One of the very useful features of \eclipse\ is that similar (sometimes identical) models can be used to solve the same problem using different solvers
(for example, constraint programming `ic' and linear programming `eplex' \cite{shen2005eplex}).
Linear (integer) programming is almost always much more effective than constraint programming (especially when looking not just for any feasible, but for the optimal solution), 
but constraint programming has more expressive power because of possible usage of non-linear constraints and objectives.
Some problems can be adequately modeled as linear (integer) programming problems, but it may not be obvious how to formulate the model in terms of linear constraints and objective --
additional variables and specific linearizing ``tricks" might be required \cite{williams2013model,aimms-model,lpsolve-abs}.
So a useful technique is to solve small input of a GCJ problem using a constraint programming model (easier to formulate, but less effective), 
and then convert the constraint programming model into a linear programming model to solve large input using linear (integer) programming solver.
Correctness of the less obvious linear programming model can be verified by comparing its results with results of the constraint programming model on the same (small) input.

Constraint logic programming model for this particular problem is easy to formulate in \eclipse\ 
(based on the fact that logic values of arithmetic constraints -- 0 or 1 -- can be used in other arithmetic constraints as integers):

\begin{lstlisting}[caption={Constraint programming solution for ``Dancing With the Googlers''}]
:- lib(ic).
:- lib(branch_and_bound).
model(S, P, Points, Triplets, GtP) :-
    length(Points, N),
    length(Triplets, N),
    ( foreach(Triplet, Triplets), 
        foreach(Point, Points), 
        fromto(0, SPrev, SCurr, S), 
        fromto(0, GtPPrev, GtPCurr, GtP), 
        param(P) do
        Triplet = [Min, Med, Max],
        Triplet :: 0..10,
        Min #=< Med, Med #=< Max,
        Max - Min #=< 2,
        Max + Med + Min #= Point,
        SCurr #= SPrev + (Max - Min #= 2),
        GtPCurr #= GtPPrev + (Max >= P ) ).
find(Triplets, GtP) :-
   flatten(Triplets, Vars),
   Cost #= -GtP,
   bb_min(labeling(Vars), Cost, 
          bb_options{strategy: dichotomic}).
\end{lstlisting}

Lines 1 and 2 load `ic' constraint programming library and a library for branch-and-bound search \cite{eclipse-library-manual}.
Lines 3--17 define constraint programming model with input parameters $S$ (number of surprising triplets), $p$, and \texttt{Points} (list of triplet sums), 
and output parameters \texttt{Triplets} (list of possible triplets values) and \texttt{GtP} (number of high triplets).
Lines 4 and 5 make \texttt{Triplets} a list of the same length as \texttt{Points}.

Lines 6--17 form an \eclipse\ loop. Lines 6--10 are loop header; \texttt{foreach} lines loop over triplets and points simultaneously, and \texttt{fromto} lines collect number of possible surprising triplets (constrained to equal S after loop end) and number of possible high triplets (returned as \texttt{GtP}). Lines 11--17 are loop body: they provide the representation of a triplet as minimum, medium and maximum elements to simplify calculation of constraint expressions for surprising and high triplets.

Lines 18--22 define the predicate \texttt{find} that finds concrete values of \texttt{Triplets} and \texttt{GtP} using \texttt{bb\_min} from the \texttt{branch\_and\_bound} library.
In line 19 \texttt{Triplets} list is flattened to \texttt{Var}, because \texttt{labeling} works only with flat lists or arrays.
\texttt{bb\_min} minimizes the objective, so in line 20 we define the objective (\texttt{Cost}) as negative of (\texttt{GtP}) that we want to maximize.

An integer linear programming model is very similar, but finds solution much faster and solves large input.

\begin{lstlisting}[caption={Integer programming solution for ``Dancing With the Googlers''}]
:- lib(eplex).
model(S, P, Points, Triplets, GtP) :-
    integers(GtP),
    length(Points, N),
    length(Triplets, N),
    ( foreach(Triplet, Triplets), 
        foreach(Point, Points), 
        fromto(0, SPrev, SCurr, S), 
        fromto(0, GtPPrev, GtPCurr, GtP), 
        param(P) do
        Triplet = [Min, Med, Max], 
        Triplet $:: 0..10, integers(Triplet),
        Min $=< Med, Med $=< Max,
        Max + Med + Min $= Point,
        Surprise $:: 0..1, integers(Surprise),
        Max - Min $=< 1 + Surprise,
        SCurr $= SPrev + Surprise,
        G $:: 0..1, integers(G),
        Max $>= G * P,
        GtPCurr $= GtPPrev + G ).
find(GtP) :-
    eplex_solver_setup(max(GtP)),
    eplex_solve(_),
    eplex_var_get(GtP, typed_solution, GtP),
    eplex_cleanup.
\end{lstlisting}

This integer linear programming model uses \texttt{eplex} library and requires two additional sets of integer variables 0..1 compared to the constraint programming model. 
\texttt{Surprise} and \texttt{G} are indicator variables local to loop iteration for ``is triplet surprising" and ``is triplet high", and their use allows to linearize the constraints. 
Unlike the \texttt{ic} library, \texttt{eplex} doesn't deduce that variables must be integer from integer domain bounds, so variables have to be declared as integers explicitly.
This solution doesn't require branch-and-bound search, because integer linear programming solver already produces the optimal solution. 

Mental steps for the \eclipse\ solution could be:

\begin{enumerate}
\item \textit{Formulate triplet representation and constraints for an individual triplet.}
\item \textit{Code constraint programming model.} 
At this point we can solve the small input and make sure that our implementation is correct.
\item \textit{Apply linearization tricks to convert non-linear constraints to integer linear.}
\item \textit{Code integer linear programming model.}
\end{enumerate}

For an imperative solution we have to notice that the lowest sum of
numbers in a high unsurprising triplet is $3p-2$ (for a triplet $p, p-1, p-1$), 
and in a high surprising triplet is $3p-4$ (for a triplet $p, p-2, p-2$), but only
if $p >= 2$ (otherwise the triplet won't be surprising). After this, we
count the triplets which are high even when unsurprising $N_{high}$,
and the triplets which can be high if they are surprising $N_{surp}$.
The answer is $N_{high} + min(N_{surp}, S)$.

Mental steps for this imperative solution could be:

\begin{enumerate}
\item \textit{Notice that unsurprising triplet is high if its sum is $\geq 3p-2$, and surprising triplet is high if its sum is $\geq 3p-4$ and $p \geq 2$.}
\item \textit{Implement iteration over triplets and calculation of $N_{high}$ and $N_{surp}$.}
\end{enumerate}

Compared to our declarative solution, the imperative solution needs fewer steps, but the first step requires some math insight into the problem.
Besides, our declarative approach provides an alternative solution for the small input which can be used to validate the solution for the large input.

\subsection*{Star Wars\footnote{Problem link: \url{http://goo.gl/DtpEQl}}}
``Star Wars" was one of the harder problems from round 2 of GCJ 2008, yet it can be almost trivially modeled and solved as a linear programming problem.

The essence of the problem statement is: you are given a set of $N$ 4-tuples of integers $x_i, y_i, z_i, p_i$. Find the minimal possible $Y$ for which exists a triplet $x, y, z$ such that 
for each original tuple $|x_i - x| + |y_i - y| + |z_i - z| \le p_iY$.

Direct translation of the problem statement to a model results in non-linear constraints,
but these can be easily converted to linear constraints using the fact that $|X| \le Max$ is equivalent to a pair of linear constraints $X \le Max$ and $-X \le Max$ \cite{lpsolve-abs}.

\begin{lstlisting}[caption={Linear programming solution for ``Star Wars''}]
:- lib(eplex).
model(Xs, Ys, Zs, Ps, X, Y, Z, P) :-
    P $>= 0,
    ( foreach(Xi, Xs), foreach(Yi, Ys), 
        foreach(Zi, Zs), foreach(Pi, Ps), 
        param(X, Y, Z, P) do 
      +(Xi - X) +(Yi - Y) +(Zi - Z) $=< Pi * P,
      +(Xi - X) +(Yi - Y) -(Zi - Z) $=< Pi * P,
      +(Xi - X) -(Yi - Y) +(Zi - Z) $=< Pi * P,
      +(Xi - X) -(Yi - Y) -(Zi - Z) $=< Pi * P,
      -(Xi - X) +(Yi - Y) +(Zi - Z) $=< Pi * P,
      -(Xi - X) +(Yi - Y) -(Zi - Z) $=< Pi * P,
      -(Xi - X) -(Yi - Y) +(Zi - Z) $=< Pi * P,
      -(Xi - X) -(Yi - Y) -(Zi - Z) $=< Pi * P ).
find(P) :-
    eplex_solver_setup(min(P)),
    eplex_solve(_),
    eplex_var_get(P, typed_solution, P),
    eplex_cleanup.
\end{lstlisting}

Line 1 loads linear programming library `eplex'.
Lines 2--15 define a linear programming model with input arguments \texttt{Xs, Ys, Zs, Ps} (lists of $x_i$, $y_i$, $z_i$, $p_i$),
and output parameter \texttt{P}. Other parameters of the \texttt{model} predicate (\texttt{X, Y, Z}) can be useful for debugging.
Line 3 constrains the value of \texttt{P} to be non-negative using \texttt{\$>=} from `eplex'. 
Lines 4--6 define a header of \eclipse\ loop: execute the loop body for each $x_i$, $y_i$, $z_i$, $p_i$, and do not treat variables \texttt{X, Y, Z, P} as local for the loop iterations.
Lines 7--14 specify the linearized problem constraints.

Lines 15--19 define the auxiliary predicate \texttt{find} that sets up the goal for \texttt{eplex} solver (minimize $P$), runs the solver, 
gets the typed value of $P$, and cleans up the solver environment for the following test cases. 

Possible steps to come up with this solution:

\begin{enumerate}
\item \textit{Express the problem constraints in linear form.}
\item \textit{Formulate linear programming model.}
\item \textit{Code linear programming model.} Trivial after the formulation.
\end{enumerate}

The first step towards a traditional algorithmic solution is to notice that we can use binary search to find the smallest possible solution $Y_s$: all $Y > Y_s$ will satisfy the constraints, and all $Y < Y_s$ will not. Thus we make a transition from searching for the smallest $Y$ to checking whether a certain $Y$ satisfies the constraints.

The constraints can be converted to linear form in the same way as in declarative solution. However, instead of solving a full linear programming problem, we do some more analysis to check whether solution exists in $O(1)$ time (see the official contest editorial\footnote{\url{http://goo.gl/ndjZhg}}). 

Possible mental steps for this solution:

\begin{enumerate}
\item \textit{Make transition from optimization problem to binary search combined with constraints feasibility check.}
\item \textit{Express the problem constraints in linear form.}
\item \textit{Perform constraints analysis to simplify feasibility check.}
\item \textit{Implement feasibility check.} 
\item \textit{Implement binary search.} 
\end{enumerate}

Linear programming solution in \eclipse\ needs fewer steps, and the steps themselves are less complex.

\subsection*{Mine Layer\footnote{Problem link: \url{http://goo.gl/6ZyZGc}}}

This problem from the GCJ World Finals 2008 turned out to be very tricky for the contestants, with the smallest success rate from the all problems of the round: 
only 42\% of the finalists who submitted answers for the large input got it right.

The problem describes a rectangular grid with odd number of rows, in which each square contains either one mine or no mines. 
Based on it we build a grid of integers: each integer is the total number of mines in corresponding square and eight adjacent squares. 
The task is: given the grid of mine counts, find the maximum possible number of mines in the middle row of the original grid.

An efficient integer programming model for ``Mine Layer'' is relatively straightforward to come up with and code in \eclipse:

\begin{lstlisting}[caption={Integer programming solution for ``Mine Layer''}]
:- lib(eplex).
model(Clues, Mines, MiddleSum) :-
  dim(Clues, [R, C]),
  dim(Mines, [R, C]),
  ( foreachelem(Mine, Mines) do
    Mine $:: 0..1,
    integers(Mine) ),
  ( multifor([I, J], 1, [R, C]), 
    param(R, C, Clues, Mines) do
      ( multifor([Di, Dj], -1, 1), 
        fromto(0, Prev, Curr, S), 
        param(I, J, R, C, Mines) do
        ( I + Di > 0, I + Di =< R, 
          J + Dj > 0, J + Dj =< C ->
          Curr = Prev + Mines[I + Di, J + Dj]
        ;
          Curr = Prev
        )
      ),
      Clues[I, J] $= S 
  ),
  ( for(J, 1, C), 
    fromto(0, Prev, Curr, MiddleSumExpr), 
    param(Mines, R) do
      Curr = Prev + Mines[R // 2 + 1, J] 
  ),
  integers(MiddleSum),
  MiddleSum $= MiddleSumExpr.
find(MiddleSum) :-
  eplex_solver_setup(max(MiddleSum)),
  eplex_solve(_),
  eplex_var_get(MiddleSum, 
    typed_solution, MiddleSum),
  eplex_cleanup.
\end{lstlisting}

Line 1 loads `eplex' library. Lines 2--29 define integer programming model with input parameter \texttt{Clues} (2-D grid)
and output parameters \texttt{Mines} (2-D array of possible mine positions, 1 if mine is present and 0 otherwise) and \texttt{MiddleSum} (count of mines in the middle row).
Line 3 gets number of rows (\texttt{R}) and columns (\texttt{C}) from the \texttt{Clues} array, and line 4 defines the same dimensions for the \texttt{Mines} array.
Lines 5--7 define domain for each element of the \texttt{Mines} as 0 or 1.

Lines 8--20 contain two nested \texttt{multifor} loops (which are syntactic sugar for several \texttt{for} loops).
For each grid square (\texttt{multifor([I, J], 1, [R, C])}) the loops construct expressions for number of mines in the square itself and in the neighboring squares,
and then constrain this expression to agree with the value in the \texttt{Clues} array.
Checks in lines 13 and 14 prevent accessing values outside of the grid for border squares.

Lines 21--27 construct expression for sum of values in the middle row of the \texttt{Mines} array (count of mines in the middle row), 
and constrain value of \texttt{MiddleSum} to be equal to value of this expression and to be integer.
It is not necessary to explicitly enforce \texttt{MiddleSum} to be integer because it is a sum of integer elements of \texttt{Mines};
but integer specification allows to output the answer directly as integer, without decimal point.

Lines 28--32 define the \texttt{find} predicate that sets up the goal for \texttt{eplex} solver (maximize \texttt{MiddleSum}), runs the solver, 
gets the typed value of \texttt{MiddleSum}, and cleans up the solver environment for the following test cases. 

Possible steps to come up with this solution:

\begin{enumerate}
\item \textit{Notice that all constraints and objective are linear.}
\item \textit{Formulate integer programming model.}
\item \textit{Code integer programming model.} The code is short, but uses a lot of \eclipse\ syntax constructs.
\end{enumerate}

\begin{table}
\centering
\caption{Running times for small (4 minutes time limit) and large (8 minutes time limit) inputs\protect\footnotemark}
\label{table:times}
\begin{tabular}{lcrr}
\hline
Problem        & Library &  Small   & Large \\
\hline
Triangle Areas & ic      & 0.2s & 1.4s \\

Dancing With 
the Googlers   & ic      & 0.2s & timeout \\

Dancing With 
the Googlers   & eplex   & 0.3s & 1.7s \\

Star Wars      & eplex   & 0.2s & 0.4s \\

Mine Layer     & eplex   & 0.2s & 2.9s \\
\hline
\end{tabular}
\end{table}
\footnotetext{Results were obtained on a 64-bit Linux machine with Intel Core i7-4900MQ CPU @ 2.80GHz and 16GB RAM using \eclipse\ 6.1 \#191.
For linear (integer) programming free COIN-OR solvers bundled with \eclipse\ were used. }

The key observation for an algorithmic solution is that we don't need to reconstruct the exact grid of mines to get the answer, 
it's enough to be able to count mines in certain sections of the grid. 
To find the number of mines in a certain section of the grid, we can split this section into $3 \times 3$ blocks of squares and add up numbers in the central squares of each block. 
If the section width or height is not divisible by 3, blocks along the sides of the original grid can have width or height of 2, 
so that the number in the ``central" square of the block still contains the total number of mines in the block. 
The solution itself is easy to implement and only requires careful handling of different sizes of the grid. 
For a full solution see the official editorial\footnote{\url{goo.gl/K2dfPi}}.

Mental steps:

\begin{enumerate}
\item \textit{Notice that instead of reconstructing the grid of mines we can count mines in certain sections of the grid.}
\item \textit{Figure out how to count mines in any $3 \times 3$ block.}
\item \textit{Figure out how to count mines in $2 \times 3$, $3 \times 2$ or $2 \times 2$ block along the border of the grid.}
\item \textit{Figure out how to split the whole grid or its parts in countable blocks various sizes of grid.}
\item \textit{Implement the counting.}
\end{enumerate}

Our declarative solution has fewer steps, and they are much easier to perform, as they don't require much insight into the problem.

\section{Conclusions}

Many GCJ problems that are hard to solve in time-restricted and stressful competition environment can be relatively easily
modeled and solved in \eclipse. 
We gave several examples of such problems, and our declarative solutions for them require simpler and often fewer mental steps than possible imperative solutions in a language like C++ or Java.
Running times of our \eclipse\ programs are several orders of magnitude smaller than the time limit imposed by GCJ rules (table \ref{table:times}).

Other modern declarative high-level tools can also be successfully used to solve competitive programming problems:
answer set programming tools, satisfiability modulo theories solvers, etc. This can be a topic of further research.

\bibliographystyle{abbrv}
\bibliography{gcj-eclipse}

\end{document}